# Spin Transport and Relaxation in Graphene

Wei Han, K. M. McCreary, K. Pi[1], W. H. Wang[2], Yan Li[3], H. Wen, J. R. Chen, R. K. Kawakami[4]

Department of Physics and Astronomy, University of California, Riverside, CA 92521

[1]Present address: Hitachi Global Storage Technologies, San Jose, CA 95135
[2]Present address: Institute of Atomic and Molecular Sciences, Academia Sinica, Taipei 106, Taiwan.
[3]Present address: Los Alamos National Laboratory, Los Alamos, NM 87545
[4]e-mail: roland.kawakami@ucr.edu

**Abstract:**

We review our recent work on spin injection, transport and relaxation in graphene. The spin injection and transport in single layer graphene (SLG) were investigated using nonlocal magnetoresistance (MR) measurements. Spin injection was performed using either transparent contacts (Co/SLG) or tunneling contacts (Co/MgO/SLG). With tunneling contacts, the nonlocal MR was increased by a factor of ~1000 and the spin injection/detection efficiency was greatly enhanced from ~1% (transparent contacts) to ~30%. Spin relaxation was investigated on graphene spin valves using nonlocal Hanle measurements. For transparent contacts, the spin lifetime was in the range of 50-100 ps. The effects of surface chemical doping showed that for spin lifetimes on the order of 100 ps, impurity scattering (Au) was not the dominant mechanism for spin relaxation. While using tunneling contacts to suppress the contact-induced spin relaxation, we observed the spin lifetimes as long as 771 ps at room temperature, 1.2 ns at 4 K in SLG, and 6.2 ns at 20 K in bilayer graphene (BLG). Furthermore, contrasting spin relaxation behaviors were observed in SLG and BLG. We found that Elliot-Yafet spin relaxation dominated in SLG at low temperatures whereas Dyakonov-Perel spin relaxation dominated in BLG at low temperatures. Gate tunable spin transport was studied using the SLG property of gate tunable conductivity and incorporating different types of contacts (transparent and tunneling contacts). Consistent with theoretical predictions, the nonlocal MR was proportional to the SLG conductivity for transparent contacts and varied inversely with the SLG conductivity for tunneling contacts. Finally, bipolar spin transport in SLG was studied and an electron-hole asymmetry was observed for SLG spin valves with transparent contacts, in which nonlocal MR was roughly independent of DC bias current for electrons, but varied significantly with DC bias current for holes. These results are very important for the use of graphene for spin-based logic and information storage applications.

# 1. Introduction

Spintronics utilizes the electron spin degree of freedom for information storage and logic operations, which could decrease the power consumption, increase data processing speed, and increase integration densities [1-3]. Lateral spin valves consisting of ferromagnetic (FM) electrodes connected to a nonmagnetic spin transport channel are of special interest because of the design flexibility for multi-terminal devices and the ability to manipulate spin during transport [4-6]. Experimentally, spin injection and transport have been observed in a variety of materials including metals, semiconductors, and carbon-based materials. The first electronic spin injection and detection was performed by Johnson and Silsbee in 1985 in a single-crystal aluminum bar at temperatures of 77 K [7]. Following this work, electron spin injection in metals, such as Al, Cu, Ag and Au, has been demonstrated even up to room temperature (RT) in some cases [5, 6, 8-13]. Spin transport in semiconductors was first detected using ultrafast optical methods [14-16]. Recently there has been significant progress in the area of electrical injection and manipulation of spin in semiconductors such as Si, GaAs, Ge, etc [17-24].

Carbon based materials have attracted considerable interest because they are expected to have long spin lifetimes due to low intrinsic spin orbit coupling and hyperfine couplings [25, 26]. Spin injection into carbon nanotubes has been performed using Co and other FM electrodes [27-33]. Organic semiconductors, such as Alq$_3$, have also been studied for spin transport due to the chemical flexibility and optoelectronic properties [34-36]. In 2004, a single atomic layer of graphitic carbon, known as graphene, was isolated by the Geim group [37]. Its carrier concentration and conductivity were found to be highly tunable with electrostatic gates, and the half-integer quantum Hall effect demonstrated the distinctive character of two-dimensional chiral Dirac fermions [38, 39]. Due to high electronic mobility, gate tunability, and the potential for long spin lifetimes, graphene has drawn a lot of attention in the spintronics field [26, 40-71]. Currently, several groups have demonstrated spin transport in single layer graphene (SLG) and multilayer graphene (MLG) [40, 42-45, 49, 50, 55]. The pioneering work was done by the van Wees group who demonstrated gate tunable spin transport and spin precession in nonlocal SLG spin valves at RT [41]. In that work, the electrical detection of spin precession was particularly important to prove that the observed signals indeed originated from spin transport. Similar results were reported by several groups [40, 42-44, 49, 55]. Subsequent results include the measurement of anisotropic spin relaxation [46], local spin transport in MLG [44, 45], spin drift effects [47], and bias dependence of spin injection [48, 50].

These earlier studies identified two critical challenges which must be overcome in order to realize the full potential of graphene for spintronics. The first important challenge was to enhance the spin injection efficiency, which was low due to the conductance mismatch between the FM metal electrodes and graphene [72]. Although it was expected that the conductance mismatch problem could be alleviated by inserting tunnel barriers into the spin injection interface [73, 74], growing smooth layers on top of graphene was non-trivial because the low surface energy and high surface diffusion led to cluster formation. It was found that using a submonolayer Ti seed layer followed by MgO deposition produced atomically smooth MgO films [75]. As a result, tunneling spin injection was achieved with greatly enhanced spin injection efficiencies [58]. The second important challenge was to determine the cause of the unexpectedly short spin lifetimes measured by the Hanle effect (spin precession) in SLG (50-200 ps) [41, 52-54, 57]. These lifetimes were orders of magnitude shorter than expected from the intrinsic spin-orbit couplings (~μs) [25, 26]. The linear scaling of spin scattering and momentum scattering in SLG [52] suggested that these were related by an Elliot-Yafet mechanism for spin relaxation (i.e. spin scattering during a momentum scattering event). Because charged impurity scattering was a strong source of momentum scattering, its effect on spin scattering was investigated through metal-doping experiments. Interestingly, it was found that for spin lifetimes on the order of 100 ps, the charged impurity scattering was not the dominant mechanism for spin relaxation in graphene [57]. Subsequently, it was observed that tunnel barriers significantly enhanced the measured spin lifetime, indicating that metal contact induced effects were very important for spin relaxation [58]. Furthermore, with tunneling contacts to suppress the contact-induced spin relaxation, the spin lifetimes observed were as long as 771 ps at RT in SLG, 1.2 ns at 4 K in SLG, 2.0 ns at RT in bilayer graphene (BLG), and 6.2 ns at 20 K in BLG [76, 77]. Very recently, Dyakonov-Perel (DP) spin relaxation was found to dominate in BLG [76-78]. Spin transport has also been demonstrated in large area epitaxial graphene fabricated by chemical vapor deposition (CVD) [79], which makes graphene a promising candidate for large scale spintronic applications [78].

In this paper, we review the contributions of our group to the study of graphene spintronics, including some of the key advances mentioned above [50, 51, 53, 57, 58, 75, 76]. This paper is constructed as follows: In section 2, we discuss the nonlocal magnetoresistance (MR) measurements. In section 3, we describe the fabrication of the graphene spin valves. In section 4, we describe our experimental results on spin injection using either transparent contacts (Co/SLG) or tunneling contacts (Co/MgO/SLG). In section 5, we investigate spin relaxation in SLG and BLG. Section 6 discusses the gate tunable spin transport in SLG for future spin field effect transistors and section 7 discusses the unique opportunities of using SLG for bipolar spintronics. In section 8, we discuss the future directions of graphene spintronics.

## 2. Nonlocal Spin Transport Measurements

Typically, there are two geometries for electrical spin transport measurements. First is the conventional spin transport geometry, known as the "local" measurement, which measures the resistance across two ferromagnetic electrodes (Figure 1a). Spin polarized electrons are injected from one electrode, transported across the graphene, and detected by the second electrode. The spin transport is detected as the difference in resistance between the parallel and anti-parallel magnetization alignments of the two electrodes. This is the geometry used for magnetic tunnel junctions and current-perpendicular-to-the-plane giant magnetoresistance (CPP-GMR) [80, 81]. The second geometry is the "nonlocal" measurement [5, 7] which uses four electrodes, as shown in Figure 1b. Here, a current source is connected across Co electrodes E1 and E2 to inject spins at E2. For spin detection, a voltage is measured across Co electrodes E3 and E4, and the signal is due to the transport of spins from E2 to E3. This measurement is called "nonlocal" because the voltage probes lie outside of the current loop. After the spin injection, the spins at E2 are able to diffuse in both directions, toward E1 (as a spin current with charge current) and toward E3 (as a spin current without charge current). This spin diffusion is usually described by a spin dependent chemical potential ($\mu_\uparrow$ and $\mu_\downarrow$), where a splitting of the chemical potential corresponds to the spin density in the graphene. Figures 1c and 1d show that as the spins diffuse toward E3, the spin density decays due to spin flip scattering. Thus, the voltage will be positive for the parallel alignment of E2 and E3 ($V_P > 0$, Figure 1c, black dots) and negative for the antiparallel alignment ($V_{AP} < 0$, Figure 1d, black dots). The signal generated by the spin transport is the nonlocal MR, defined as $\Delta R_{NL} = (V_P - V_{AP})/I$, where $I$ is the injection current. Experimentally, there is usually a nonlocal baseline resistance which could be due to leakage current, Peltier and Seebeck effects, and nonconserving spin scattering at the interface [82-84].

Comparing these two geometries, the nonlocal measurement is more sensitive to detect the spin signal because the spin current is isolated from the charge current. As discussed later, the nonlocal MR measurement has much better signal-to-noise ratio compared to the local MR measurement performed on the same device. Also, the nonlocal measurement is less prone to artifacts such as anomalous Hall and anisotropic magnetoresistance (AMR) effects [5].

In our study, the electrical measurements were performed using lock-in detection with frequencies in the range of 7-17 Hz. The ac current excitation was ~ 1 µA for spin valves with tunneling contacts and ~ 30 µA for spin valves with transparent contacts.

## 3. Device Fabrication

The graphene spin valves (Figure 2a, $L$ is the spacing between injector and detector) were fabricated as follows. First, graphene sheets were exfoliated from highly oriented pyrolytic graphite (HOPG, grade ZYA, SPI supplies) and placed on 300 nm $SiO_2$/Si(001) substrates [37, 85], where the Si wafer was degenerately doped and used as a back gate. The typical size of the graphene was between 1 and 6 µm wide and up to 50 µm long. The SLG and BLG were identified by optical microscopy and the thickness was verified by Raman spectroscopy [86]. Figure 2b shows typical spectra from SLG and BLG measured on our devices. The Co electrodes were defined by electron-beam lithography and fabricated using angle evaporation through the patterned PMMA/MMA bilayer resist, which had a slight undercut (The thickness of MMA was ~400 nm, and the thickness of PMMA was ~ 200 nm). The electrodes were deposited in a molecular beam epitaxy (MBE) system with a base pressure ~ $2\times10^{-10}$ torr. No cleaning procedure was used after the resist development and before the angle deposition in MBE. MgO was deposited from a single crystal source by electron-beam evaporation. Co was deposited by thermal evaporation and Ti was deposited by electron-beam evaporation. The typical growth rates of these materials were ~1 Å/min. The temperature of the sample was less than 40 °C during growth with a water-cooled sample stage to prevent annealing of the MMA and PMMA films.

For SLG spin valves with transparent contacts, the following growth procedure was used. First, a 2 nm MgO masking layer was deposited with an angle of 0° from normal incidence. Then an 80 nm Co layer was deposited with an angle of 7° (figure 2d). This angle evaporation reduced the width of the contact area to ~50 nm (=400 nm × tan 7°), which was expected to increase the spin signal [87]. The spin valves with transparent contacts had a yield of 70%, defined as the percentage of electrodes that exhibit spin transport.

For SLG spin valves with tunneling contacts, they were grown as follows. First, 0.12 nm of Ti was deposited at both 0° and 9° angles, followed by oxidation in $5\times10^{-8}$ torr of $O_2$ for 30 minutes to convert the metallic Ti into insulating $TiO_2$, which promoted the growth of ultrathin atomically smooth MgO films (This will be discussed in detail later) [75]. Then 3 nm of MgO was deposited at an angle of 0° for the masking layer and 0.8 nm of MgO was deposited at an angle of 9° for the tunnel barrier. Then the 80 nm thick Co electrode was deposited with an angle of 7° (figure 2e). The yield of these tunneling contacts were ~ 5%, defined as the percentage of electrodes that exhibit spin transport and tunneling characteristics.

Later, in the spin relaxation study with tunneling contacts, we used Au contacts as the outer two electrodes, instead of Co, in order to achieve symmetric Hanle curves [76]. A first step of electron beam lithography was used for the Au contacts. Then, the device was annealed on a heater at 150 °C (heater temperature) for 1 hour in vacuum (~ $3\times10^{-9}$ torr) to clean the graphene surface. Then a second electron beam lithography step was used for the Co contacts, which follows the same fabrication procedure for tunneling contacts as discussed earlier. It was found that the yield of the tunneling contacts increased to ~30% when the Au electrode fabrication and surface cleaning in vacuum were added. For all devices, no extra cleaning or annealing procedures were done after the fabrication.

Prior to lift-off, graphene spin valves with either transparent contacts or tunneling contacts were quickly loaded in an electron beam evaporator (~ $5\times10^{-6}$ torr) and capped with 5 nm $Al_2O_3$ to protect the Co from further oxidation. The lift-off was done in PG remover on a hot plate with a temperature of 70 °C. Figure 2c shows a scanning electron microscope image of a SLG spin valve device, in which the darker region corresponds to the SLG, and figure 2f shows a typical optical image of a graphene spin valve The widths of the Co electrodes in our studies were varied between 90 nm and 500 nm to have different coercivities. The 90° turn in the Co electrode (in the red dashed circle) was used to inhibit domain wall motion to help generate the antiparallel magnetization alignment in the magnetic field sweeps.

After fabrication, the graphene spin valves were stored in UHV with a pressure of ~ $1\times10^{-9}$ torr to protect the Co from oxidation. The carrier concentration of SLG and BLG were tuned by the back gate voltage. A typical gate dependence of SLG conductivity is shown in figure 2g. The peak of the resistance corresponds to the Dirac point ("charge neutrality point", or CNP, for BLG). In our study, the mobility of SLG was in the range of 1000-5000 $cm^2$/Vs and the mobility of the BLG was in the range of 400-2000 $cm^2$/Vs.

## 4. Spin Injection

For lateral spin valves, there are three important processes: spin-polarized current generation, spin transport, and spin detection. Graphene is an interesting material for lateral spin valves due to gate tunable conductivity [37, 38, 85], spin transport at RT, and long spin diffusion lengths (several μm at RT) [40-44, 49, 55]. In the following, we discuss our work on spin injection using transparent contacts (Co/SLG) and tunneling contacts (Co/MgO/SLG).

**4.1 Spin injection using transparent contacts**

The geometry of the Co/SLG contact is shown in Figure 3a, where the Co is directly contacted to SLG with a 2 nm MgO masking layer. The linear characteristic of differential contact resistance indicates the ohmic contact behavior (Figure 3b). The typical nonlocal MR curve is shown in figure 3c (measured on device A with transparent contacts at RT with $V_g = 0$ V). As the magnetic field is swept up from negative to positive, one electrode, E2, switches the magnetization direction first, resulting in the negative nonlocal MR value due to the changing from parallel to anti-parallel state. Then, the electrode E3 switches the magnetization direction, and the state changes back to the parallel state. The minor jumps in the nonlocal MR loop are from the switching of the electrodes E1 and E4. The arrows in figure 3c indicate the magnetic directions of the four Co electrodes.

Using a single SLG sheet contacted by seven Co electrodes at various spacings (Device B, transparent contacts, as shown in figure 4a), we investigated the dependence of the spin transport and spin precession as a function of distance. Figures 4b-4d show the nonlocal MR loop for 1 μm, 2 μm, and 3 μm spacings, respectively, measured at 300 K with $V_g = 0$ V. The nonlocal MR decreases from 100 mΩ to 2 mΩ as the spacing increases from 1 μm to 3 μm. The spin injection efficiency is calculated to be ~ 1 %, based on the following equation:

$$\Delta R_{NL} = \frac{1}{\sigma_G} \frac{P^2 \lambda_G}{W} e^{-L/\lambda_G} \quad (1)$$

where $P$ is the spin injection/detection efficiency, and $\sigma_G$, $W$ and $\lambda_G$ are the conductivity, width, and spin diffusion length of the SLG, respectively. This low spin injection efficiency of 1% is expected due to the conductance mismatch between Co and SLG [72].

### 4.2 Tunnel barrier growth

For spin injection into semiconductors, tunnel barriers have been used to alleviate the conductance mismatch problem [17, 19, 21, 73, 74, 88]. However, for graphene, the growth of uniform tunnel barriers on the surface is non-trivial. As shown in Figure 5b, the growth of 1 nm MgO on top of HOPG tends to cluster and form "pinholes" due to the high surface diffusion and low surface energy of HOPG/graphene. Similar problems were also observed by researchers growing $HfO_2$ or $Al_2O_3$ on graphene [89-91]. After trying different techniques, we developed a method for growing atomically smooth MgO films on graphene by using a submonolayer Ti seed layer. Figure 5a shows the schematic drawing of the MgO grown on top of the Ti seed layer. With only 0.5 ML Ti (0.12 nm) prior to the MgO growth, the RMS roughness of the MgO film dropped quickly from 0.7 nm, for MgO on bare HOPG, to 0.2 nm (figure 5b and 5c) [75]. Figure 5d (open circles) shows the RMS roughness of 1 nm MgO films on HOPG as a function of the thickness of Ti. It is clearly shown that the smoothness of the film greatly improves as the Ti thickness increases from 0 ML to 0.5 ML. To use this as a tunnel barrier, we oxidized the Ti in order to form $TiO_2$ to avoid a possible conductive path between Co and SLG. Finally, we applied this technique to SLG flakes, and found the RMS roughness of a 1 nm MgO film is less than 0.2 nm (figure 5e and 5f).

### 4.3 Spin injection using tunneling contacts

Using the Ti-seeded MgO barrier, we fabricated SLG spin valves with tunneling contacts (figure 6a). The current-voltage (*I-V*) curves between electrodes were highly non-linear (figure 6b), and differential contact resistance showed a very sharp peak at $I_{dc} = 0$ μA (figure 6c), which is an indication of tunneling between the Co electrodes and SLG. A 130 Ω nonlocal MR was observed for a SLG spin valve measured at 300 K with $V_g = 0$ V (Device C, tunneling contacts).

The spacing between injector and detector ($L$) ~ 2.1 µm, and the width of SLG ($W$) ~ 2.2 um, as shown in figure 6d. The spin injection efficiency, $P$, was calculated to be 26%-30% using equation 1 with experimental values of $\sigma_G$ = 0.35 mS, and typical experimental values of $\lambda_G$ = 2.5-3.0 µm. This high spin injection/detection efficiency highlighted the high quality of the MgO tunnel barrier, which alleviated the conductance mismatch between Co and SLG in spin valves having transparent contacts [73, 74]. This compares favorably with the tunneling spin polarization of 35% - 42% measured by spin-dependent tunneling from Co into a superconductor across polycrystalline $Al_2O_3$ barriers [80, 92, 93]. Figure 6e and figure 6f show the nonlocal and local MR loops measured at 4 K (Device C, tunneling contacts). The nonlocal MR and local MR signal were ~ 100 Ω and ~ 200 Ω, respectively. This local MR is roughly twice the nonlocal MR, which is precisely the behavior expected theoretically [94, 95]. Also, the nonlocal MR loop has much better signal-to-noise ratio and is more sensitive to detect the spin signal compared to the local MR.

## 5. Spin Relaxation

SLG is a promising material for spintronics because it is predicted to have long spin relaxation times and long spin diffusion length due to the low intrinsic spin-orbit and hyperfine couplings [25, 26, 68, 69, 96]. However, the measured spin lifetimes in SLG (50-200 ps) were orders of magnitude shorter than expected from the intrinsic spin-orbit coupling [25, 41, 51, 52, 54, 55, 57, 97]. Here, we utilized Hanle spin precession measurements to investigate the role of charged impurity scattering and contact induced effects on the spin relaxation in graphene. Furthermore, we observed long spin lifetimes in SLG and BLG.

### 5.1 Hanle spin precession

Hanle spin precession was performed by applying an out-of-plane magnetic field, with the geometry shown in figure 7a. The magnetic field induced spin precession at a Larmor frequency of $\omega_L = g\mu_B H_\perp/\hbar$, where $g$ is the g-factor, $\mu_B$ is the Bohr magneton, and $\hbar$ is the reduced Planck's constant. Figure 7b shows typical Hanle spin precession curves, which were obtained by measuring the nonlocal resistance as a function of $H_\perp$ for a SLG spin valve with transparent contacts at 300 K (Device D, $L$ = 3 μm). The top branch (red curve) is for the parallel magnetization state of the central electrodes, and the bottom branch (black curves) is for the antiparallel magnetization state. The characteristic reduction in the spin signal with increasing magnitude of $H_\perp$ was a result of spin-precession induced by the out-of-plane field, which reduces the spin polarization reaching the detector electrode. Quantitatively, the Hanle curve depends on spin precession, spin diffusion, and spin relaxation and is given by:

$$R_{NL} \propto \pm \int_0^\infty \frac{1}{\sqrt{4\pi Dt}} \exp\left[-\frac{L^2}{4Dt}\right] \cos(\omega_L t) \exp(-t/\tau_s) dt \qquad (2)$$

where the + (-) sign is for the parallel (antiparallel) magnetization state, $D$ is the diffusion constant, and $\tau_s$ is the spin lifetime [6]. Using this equation, we fitted the parallel and anti-parallel Hanle curves (solid lines). The fitting parameters obtained were $D = 2.5\times10^{-2}$ m$^2$s$^{-1}$ and $\tau_s$ = 84 ps, which corresponded to a spin diffusion length of $\lambda_s = \sqrt{D\tau_s}$ = 1.5 μm. This Hanle lifetime represents a lower bound of spin lifetime due to contact induced spin relaxation, which will be discussed in detail later.

**5.2 Charged impurity scattering**

Our first study was to systematically introduce additional sources of charged impurity scattering and monitor their effect on spin lifetime [57]. Gold impurities were selected for this purpose because they were shown to behave as charged impurity scatterers with 1/$r$ Coulomb potential when deposited at low temperature without clustering (18 K) [98] and were not expected to generate other effects such as resonant scattering, wavefunction hybridization, or chemical bonding. Figure 8a shows the effect of Au doping (at 18 K) on the conductivity of SLG as a function of gate voltage (Device E, transparent contacts). The mobility decreased and the Dirac point shifted to more negative gate voltages as the doping increased. Figure 8b shows the best fit values of $\tau_s$ and D, achieved by fitting Hanle data for device E, as a function of Au coverage at the Dirac point (black squares), for an electron concentration of $2.9\times10^{12}$ cm$^{-2}$ (red circles), and for a hole concentration of $2.9\times10^{12}$ cm$^{-2}$ (blue triangles). In all three cases, $\tau_s$ did not decrease with increasing Au coverage while the corresponding values of $D$ decreased as a function of Au coverage (figure 8b inset). These results clearly show that for spin lifetimes on the order of 100 ps, charged impurity scattering is not the primary source of spin relaxation in graphene, even though it is very effective at generating momentum scattering.

**5.3 Contact induced spin relaxation**

Measurement of Hanle spin precession in SLG spin valves having transparent, pinhole, and tunneling contacts shows that contact induced effects are very important to the spin relaxation in SLG. For SLG spin valves with tunneling contacts, the spin lifetimes measured at the Dirac point were 771 ps and 448 ps with 4.5 μm (Device F) and 5.5 μm spacing (Device G), respectively at 300 K, as indicated in figure 9a and 9b. These are much longer than the spin lifetime of 134 ps measured for pinhole contacts (Device H, figure 9c) and 84 ps for transparent contacts (figure 7b) at 300 K, which are consistent with the values reported in previous studies (50 - 200 ps) [41, 51, 52, 54]. Due to the increased spin lifetimes, the spin diffusion lengths from the Hanle fits ($\lambda_G = \sqrt{D\tau_s}$) are significantly larger for tunneling contacts (2.5-3.0 μm) than for transparent and pinhole contacts (1.2-1.4 μm). The longer spin lifetimes and spin diffusion lengths with tunneling contacts indicate that the effect of the contact-induced relaxation is substantial for transparent and pinhole contacts.

The ferromagnetic contacts can theoretically introduce spin relaxation through a number of mechanisms. One mechanism is through the inhomogeneous magnetic fringe fields. The non uniform interface could cause the local Co electrode magnetization to be inhomogeneous and non-collinear with the average magnetization, which will generate spin relaxation through spin precession about random local fields. This mechanism has been proposed to be significant in semiconductor systems [99]. A second mechanism is interfacial spin scattering, which is possible because of the direct contact between the FM and graphene for the transparent and pinhole contacts. A third mechanism for contact-induced spin relaxation is related to the Hanle measurement itself. With a metallic Co in contact with graphene, the spins diffuse from the graphene to the Co with a characteristic escape time, $\tau_{esc}$. Due to the conductance mismatch between Co and graphene, the escape time can become comparable to or less than the actual spin lifetime (denoted as the spin-flip time, $\tau_{sf}$). In this case, the Hanle lifetime ($\tau_s$) in equation 2 is determined by both the spin-flip time and the escape time, with a simple relationship of $\tau_s^{-1} = \tau_{sf}^{-1} + \tau_{esc}^{-1}$ when the spin diffusion length is much larger than the sample size [100]. Because the Hanle lifetime is determined mostly by the smaller of $\tau_{sf}$ and $\tau_{esc}$, it can only provide a lower bound of the spin lifetime. More accurate measurement of the true spin lifetime in graphene is made possible by the insertion of good tunnel barriers to reduce the out-diffusion of spins. This greatly increases $\tau_{esc}$ to yield $\tau_s^{-1} \approx \tau_{sf}^{-1}$ when $\tau_{esc} \gg \tau_{sf}$, so that the measured Hanle lifetime $\tau_s$ more accurately measures the actual spin-flip time in graphene $\tau_{sf}$. We note that it is possible to model this type of contact-induced spin relaxation numerically (see appendix of [54]) but an analytic expression for the Hanle curves including escape time effects is currently unavailable.

Based on the Hanle data in figures 7 and 9 with much longer lifetimes for tunneling contacts compared to pinhole and transparent contacts, it is likely that for spin valves with lifetimes in the 50-200 ps range, the dominant spin relaxation is generated by the contacts. Thus, future studies of spin relaxation will require the use of tunneling contacts to suppress the contact-induced spin relaxation.

**5.4 Spin relaxation in SLG**

Using the tunnel barrier devices to suppress the contact-induced spin relaxation, we systematically studied spin relaxation in SLG spin valves (typical device: Device G, $L = 5.5$ µm). At RT (300 K), $\tau_s$ was in the range of 400-600 ps, and there was no obvious correlation between $\tau_s$ and $D$ (Figure 10a). Interestingly, when the device was cooled to $T = 4$ K (Figure 10b), $\tau_s$ and $D$ exhibited a strong correlation, with both quantities increasing with carrier concentration. The correlation of $\tau_s$ and $D$ implies a linear relation between $\tau_s$ and the momentum scattering time, $\tau_p$ ($D \sim \tau_p$ as discussed in refs. [52, 97, 101]). This indicates that at low temperatures the spin scattering is dominated by momentum scattering through the EY mechanism (i.e. finite probability of a spin-flip during a momentum scattering event) [68, 102, 103]. The temperature dependences of $\tau_s$ and $D$ at different carrier concentrations are shown in Figures 10c and 10d. As the temperature decreases from 300 K to 4 K, $\tau_s$ shows a modest increase at higher carrier densities (e.g. from ~0.5 ns to ~1 ns for $V_g - V_{CNP} = +60$ V) and little variation for lower carrier densities. The temperature dependence of $D$ shows a similar behavior as $\tau_s$. To analyze the relationship between the spin scattering and momentum scattering, we plotted $\tau_s$ vs. $D$ for temperatures below or equal to 100 K (Figure 10e) and for temperatures above 100 K (Figure 10e inset). The main trend is that for lower temperatures, $\tau_s$ scales linearly with $D$, which indicates that an EY spin relaxation mechanism is dominant at lower temperatures ($\leq$100 K). For higher temperatures, $\tau_s$ and $D$ do not follow the linear relationship as shown at low temperatures, which suggests that multiple sources of spin scattering are present.

## 5.5 Spin relaxation in BLG

BLG differs from SLG not just in thickness but also in band structure (linear for SLG, massless fermions vs. hyperbolic for BLG, massive fermions) and intrinsic spin-orbit coupling [104, 105]. Figures 11a-c show the gate voltage dependence of $\tau_s$ and $D$ for BLG spin valve (Device I, tunneling contacts, $L = 3.1$ μm) at 20 K. The measured spin lifetimes are 2.5 ns, 6.2 ns, and 3.3 ns for $V_g$ - $V_{CNP}$ = -40, 0, 40 V. It is noted that longer spin lifetimes are observed in BLG (up to 6.2 ns) than in SLG (up to 1.0 ns). Theoretically, the intrinsic spin-orbit coupling in BLG is an order of magnitude larger than in SLG, which is predicted to result in shorter spin lifetimes for BLG [105]. The opposite experimental trend verifies that the spin relaxation in graphene is of extrinsic origin in the SLG.

It is also observed that the spin lifetimes at 20 K varies differently from RT results, where $\tau_s$ varies from 250 ps to 350 ps as a function of gate voltage and exhibits no obvious correlation with $D$ (Device I, tunneling contacts, figure 11d). A detailed measurement of the gate voltage dependence at 4 K (BLG device J, tunneling contacts, L = 4 μm) is shown in figure 11e. $\tau_s$ varies from 1 ns to 2.7 ns, showing a peak at the charge neutrality point. On the other hand, the gate voltage dependence of $D$ exhibits lower values near the charge neutrality point and increasing values at higher carrier densities. The opposite behaviors of $\tau_s$ and $D$ suggest the importance of DP spin relaxation (i.e. spin relaxation via precession in internal spin-orbit fields) where $\tau_s$ scales inversely with $\tau_p$ [103, 106].

Figures 11f and 11g show the temperature dependences of $\tau_s$ and $D$, respectively (BLG device I). At low temperatures, $\tau_s$ is enhanced while $D$ is reduced, which is different from SLG where both D and τ_s increase as temperature decreases for most gate voltages. The opposite trends of the temperature dependences of $\tau_s$ and D suggest the strong contributions of spin relaxation mechanisms of the DP type, which is also suggested in ref. [77] Possible sources of extrinsic EY spin relaxation include long-range (Coulomb) impurity scattering and short-range impurity scattering [67], while an extrinsic DP spin relaxation could arise from curvature of the graphene film (for example, random spin-orbit couplings) [25, 68, 107]. The transition from EY-dominated SLG to the DP-dominated BLG could be due to a strong reduction of the EY contribution because of enhanced screening of the impurity potential in thicker graphene [97, 108] and the smaller surface-to-volume ratio.

## 6. Gate tunable spin transport

The SLG conductivity is tunable by applying a gate voltage, which could be used for gate manipulation of spin transport. Here, taking advantage of the success of spin injection with either transparent contacts (Co/SLG), or tunneling contacts (Co/MgO/SLG), we investigated gate tunable spin transport in SLG spin valves [51, 58]. Figure 12a and 12b summarize the nonlocal MR (black squares) and SLG conductivity (red line) as a function of gate voltage for SLG spin valves with transparent contacts (Device K, $L = 1$ μm) and tunneling contacts (Device L, $L = 2.2$ μm), respectively at 300 K. It is clear that the nonlocal MR is larger at higher carrier densities for transparent contacts, while it is smaller at higher carrier densities for tunneling contacts.

To understand the observed behaviors, the nonlocal MR of SLG spin valves is derived following the method of Takahashi and Maekawa [109]:

$$\Delta R_{NL} = 4R_G e^{-L/\lambda_G} \left( \frac{P_J \frac{R_J}{R_G}}{1-P_J^2} + \frac{P_F \frac{R_F}{R_G}}{1-P_F^2} \right)^2 \times \left( \left( 1 + \frac{2\frac{R_J}{R_G}}{1-P_J^2} + \frac{2\frac{R_F}{R_G}}{1-P_F^2} \right)^2 - e^{-2L/\lambda_G} \right)^{-1} \quad (3)$$

where $R_G = 1/\sigma_G (\lambda_G/W)$, $R_F = \rho_F \lambda_F / A_J$ are the spin resistances of the SLG and FM electrodes, respectively, $W$ is the width of the SLG, $A_J$ is the junction area between the FM and SLG, $\lambda_G$ ($\lambda_F$) is the spin diffusion length in the SLG (FM), $\sigma_G$ is the conductivity of SLG, $\rho_F$ is the resistivity of the FM, $P_F$ is the spin polarization of the FM, $P_J$ is the polarization of the interfacial current, $R_J$ is the contact resistance between FM and SLG, and $L$ is the spacing between the injector and detector electrodes.

For transparent contacts ($R_J \ll R_G$), equation 3 reduces to:

$$\Delta R_{NL} = 4 \frac{1}{R_G} \left( \frac{P_J R_J}{1-P_J^2} + \frac{P_F R_F}{1-P_F^2} \right)^2 \frac{e^{-L/\lambda_G}}{1-e^{-2L/\lambda_G}} \sim \sigma_G \quad (4)$$

Hence, the nonlocal MR is roughly proportional to the conductivity of SLG.

For tunneling contacts ($R_J \gg R_G$), equation 3 reduces to:

$$\Delta R_{NL} = \frac{1}{\sigma_G} \frac{P_J^2 \lambda_G}{W} e^{-L/\lambda_G} \sim \frac{1}{\sigma_G} \quad (5)$$

Hence, the nonlocal MR varies inversely with the conductivity of SLG.

To illustrate this more clearly, we modeled the nonlocal MR behavior as a function of gate voltage. The conductivity is assumed to vary linearly with $V_g$ away from the Dirac point according to $\sigma_G = \sigma_0 + \mu e \alpha |V_g|$, where $\sigma_0$ is the minimum conductivity (assumed to be $4e^2/h$) [85], $\mu$ is the mobility (taken to be 2000 cm$^2$/Vs), $e$ is the electron charge, and $\alpha$ is the capacitance per area (taken to be $1.15 \times 10^{-8}$ F/cm$^2$ for 300 nm of SiO$_2$). For transparent contacts, due to the proportionality of $\Delta R_{NL}$ and $\sigma_G$, the gate dependence of $\Delta R_{NL}$ has a minimum at the Dirac point (figure 12c top curve). The low value of $\Delta R_{NL}$ (figure 12a) in the calculation is due to the conductance mismatch term $(R_F/R_G)^2 \ll 1$ in equation 4. Intuitively, the increase of nonlocal MR with increasing conductivity occurs because the conductance mismatch between the Co and SLG is reduced [72]. For SLG spin valves with tunneling contacts, the nonlocal MR has a maximum at the Dirac point and is inversely proportional to the SLG conductivity as shown in figure 12c bottom curve. Fundamentally, the inverse scaling of $\Delta R_{NL}$ with $\sigma_G$ is associated with the spin injection process. Specifically, spin injection produces a difference in the spin-dependent chemical potential given by $\Delta \mu = \mu_\uparrow - \mu_\downarrow = P_J R_G I/(-e)$ at the injection point under the injector. Thus, a larger $R_G$ will lead to a greater difference in the spin-dependent chemical potential, leading to a larger nonlocal MR signal.

## 7. Bipolar spintronics

Bipolar spintronics uses carriers of both polarities (electrons and holes) contributing to spin-charge coupling [110, 111]. Compared to conventional semiconductors (GaAs, Si, etc), SLG is very unique due to the symmetric band structure [112].

To use SLG for bipolar spin transport, we studied the nonlocal electron and hole spin transport under different dc bias current: $I_{dc}$ = +300 µA (squares), 0 µA (circles), and -300 µA (triangles) in device M (SLG, $V_D$ = -34 V, $L$ = 1 µm) at 300 K as shown in figure 13c. For positive bias, which means that the current is going from Co to SLG (illustrated in figure 13a), the gate-dependence of $\Delta R_{NL}$ follows the zero bias data. On the other hand, when the bias is negative and the carriers are holes (triangles, $V_g < V_D$), a strong reduction of $\Delta R_{NL}$ was observed. In this case, the holes in the SLG are driven toward the Co electrode E2 and become spin-polarized due to spin-dependent reflection from the ferromagnetic interface (i.e. spin extraction [113], as shown in figure 13b). A very interesting aspect is that the reduction of $\Delta R_{NL}$ was observed for spin extraction of holes, but not for the spin extraction of electrons.

To further study this reduction of the $\Delta R_{NL}$, we measured the nonlocal MR as a function of dc bias current at fixed gate voltages: $V_g = 0$ V (electrons, solid squares) and for $V_g = -70$ V (holes, open squares) at 300 K, as shown in figure 13d. For electrons, there is only a slight variation in $\Delta R_{NL}$ as a function of $I_{dc}$. For holes at positive bias, the behavior of $\Delta R_{NL}$ is similar to the electron case. For holes at negative bias, however, there is a significantly stronger variation of $\Delta R_{NL}$ as a function of dc current bias, with decreasing $\Delta R_{NL}$ at larger negative biases. The images in figure 13e show the $\Delta R_{NL}$ measured at 300 K as a function of both gate voltage (y) and dc current bias (x). The two main trends, namely the roughly constant $\Delta R_{NL}$ vs. $I_{dc}$ for electrons and the reduced $\Delta R_{NL}$ for hole spin extraction, can be clearly seen.

These behaviors have been reproduced on all our devices. In one particular SLG device (Device N, transparent contacts, $V_D = -32$ V, $L = 1$ μm), there was a very strong change of $\Delta R_{NL}$ as a function of dc current bias for hole doping ($V_g = -50$ V). By increasing the negative dc bias current to -660 μA, a sign reversal of $\Delta R_{NL}$ at approximately 450 μA was observed (Figure 13f).

This type of reduction and/or inversion of the spin signal as a function of dc bias has been observed in many different systems including Fe/n-GaAs [17], CoFe/Al$_2$O$_3$/Al [11], and Fe/Al$_2$O$_3$/Si [22]. In all cases, the interface includes a barrier and the reduction/inversion is observed for spin extraction as opposed to spin injection. The interesting aspect of SLG is that this effect is seen for holes but not for electrons, despite having symmetric electron and hole bands. Therefore, SLG provides a unique system to investigate this phenomenon.

## 8. Future work

Future work in graphene spintronics can proceed in several interesting directions. First of all, experimental studies of spin relaxation are needed to investigate various mechanisms for spin scattering proposed theoretically (curvature, localized magnetic moments, hydrogen doping, etc. [65-68]). Second, further enhancement of spin injection could be achieved using half metallic materials as the spin injector, such as LSMO, and $Fe_3O_4$, etc [114, 115]. Third, graphene nanoribbons are predicted to become half-metallic (100% spin polarized) with the application of a transverse electric field [59]. Spin transport in nanoribbons is also expected to generate extremely large MR ratios of over $10^6$ [60]. Fourth, magnetic doping of graphene is expected to generate interesting spin-dependent properties [116]. Fifth, there are proposals which involve using graphene quantum dots for spin qubits [26], and the favorable properties of graphene spin transport make it an ideal candidate for spin-based logic [117, 118].

## 9. Conclusion

Spin injection and transport in SLG were successfully achieved using either transparent contacts with Co electrodes directly contacted to SLG or tunneling contacts with Co electrodes separated from SLG by ultrathin MgO insulating barriers. Using nonlocal MR measurements, we were able to study the spin dependent properties in SLG. With tunneling contacts, the nonlocal MR was increased by a factor of ~1000 and the spin injection/detection efficiency was greatly enhanced to 26-30%. Spin relaxation was investigated with SLG spin valves using Hanle measurements. The effects of surface chemical doping on spin relaxation showed that for spin lifetimes on the order of 100 ps, charged impurity scattering is not the dominant mechanism for spin relaxation in graphene, despite its importance for momentum scattering. Enhanced spin lifetimes for tunneling contacts indicated that contact-induced spin relaxation was substantial.

Using tunneling contacts to suppress the contact-induced spin relaxation, we observed the spin lifetimes as long as 771 ps at RT in SLG, 1.2 ns at 4 K in SLG, and 6.2 ns at 20 K in BLG. Furthermore, contrasting spin relaxation behaviors were observed in SLG and BLG. Using the SLG property of gate tunable conductivity and incorporating different types of contacts, we studied the nonlocal MR as a function of gate voltage for both types of contacts. Consistent with theoretical predictions, the nonlocal MR was proportional to the SLG conductivity for transparent contacts and varied inversely with the SLG conductivity for tunneling contacts. An electron-hole asymmetry was discovered for SLG spin valves with transparent contacts, in which nonlocal MR was roughly independent of dc bias current for electrons, but varied significantly with dc bias current for holes. These results are important for the development of graphene spintronics for future spin-FETs, bipolar spintronics devices, and non-volatile spin logic devices.

**Acknowledgements**

We acknowledge the support of ONR (N00014-09-1-0117), NSF (CAREER DMR-0450037), NSF (DMR-1007057), and NSF (MRSEC DMR-0820414).

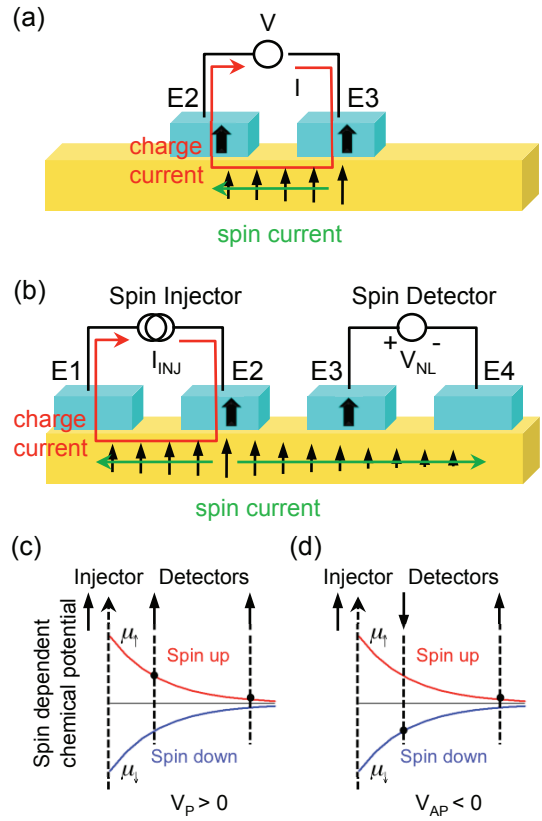

Figure 1. The spin transport measurement. (a) schematic of local spin transport measurements. (b) schematic of nonlocal spin transport measurements. (c-d) spin dependent chemical potential for parallel and anti-parallel states of spin injector and detectors in the nonlocal geometry.

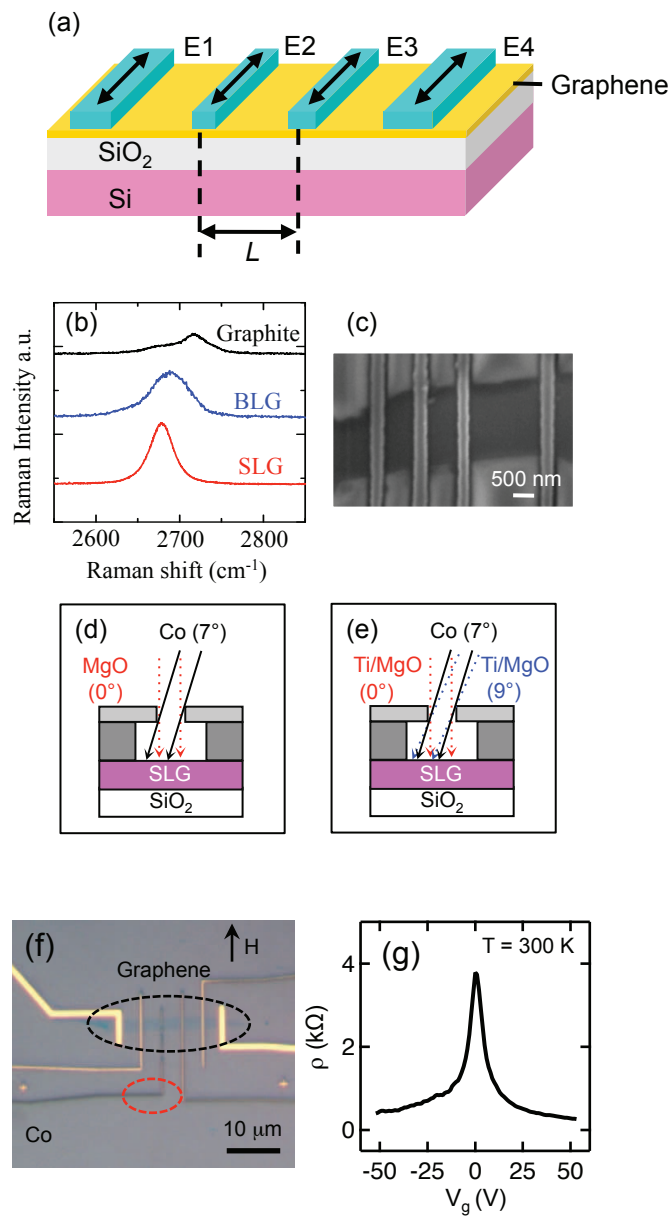

Figure 2. The graphene spin valve device structure. (a) the schematic graphene spin valves on $SiO_2$/Si substrate. (b) Raman microscopy of SLG, BLG, and graphite. (c) SEM image of a SLG spin valve device, the dark region corresponds to the SLG. (d) Angle evaporation for transparent contacts. (e) Angle evaporation for tunneling contacts. Ti is oxidized to $TiO_2$ right before MgO growth. (f) Optical image of a graphene spin valve. The magnetic field is applied along the easy axis of the Co electrodes. The 90° turn in the Co electrode (in the red dashed circle) is used to inhibit domain wall motion to help generate the antiparallel magnetization alignment in the magnetic field sweeps. (g) Typical gate voltage dependence of the resistivity of SLG.

**Figure 3**

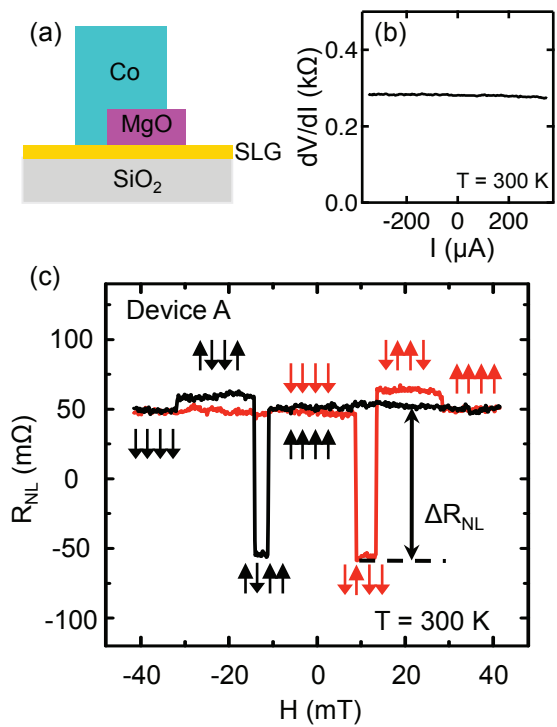

Figure 3. The SLG spin valve device with transparent contacts Co/SLG (Device A: SLG, transparent contacts, $L \sim 1$ μm, $W \sim 2$ μm). (a) the schematic diagram of the transparent contacts with 2 nm MgO masking layer. (b) Differential resistance vs. current bias at zero gate voltage at 300 K. (c) Typical non-local MR loop (measured on device A at 300 K, $V_g = 0$ V) as the magnetic field is swept up (red curve) and swept down (black curve). The arrows show the magnetization of the four Co electrodes.

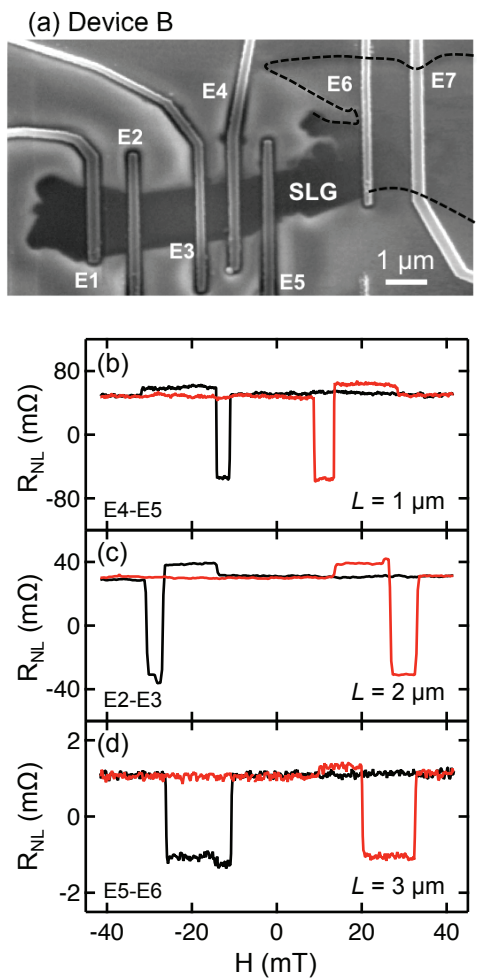

Figure 4. The SLG spin valve device with seven Co electrodes with various spacings (Device B, SLG, transparent contacts). (a) SEM image of this device. E1, E2, E3, E4, E5, E6, E7 are seven Co electrodes. Dashed lines show the edge of the SLG in a region of the image which has poor contrast. (b-d) Non-local MR scans for $L$ = 1 μm (injector: E4, detector: E5), 2 μm (injector: E2, detector: E3), 3 μm (injector: E5, detector: E6), respectively measured at 300 K ($V_g$ = 0 V).

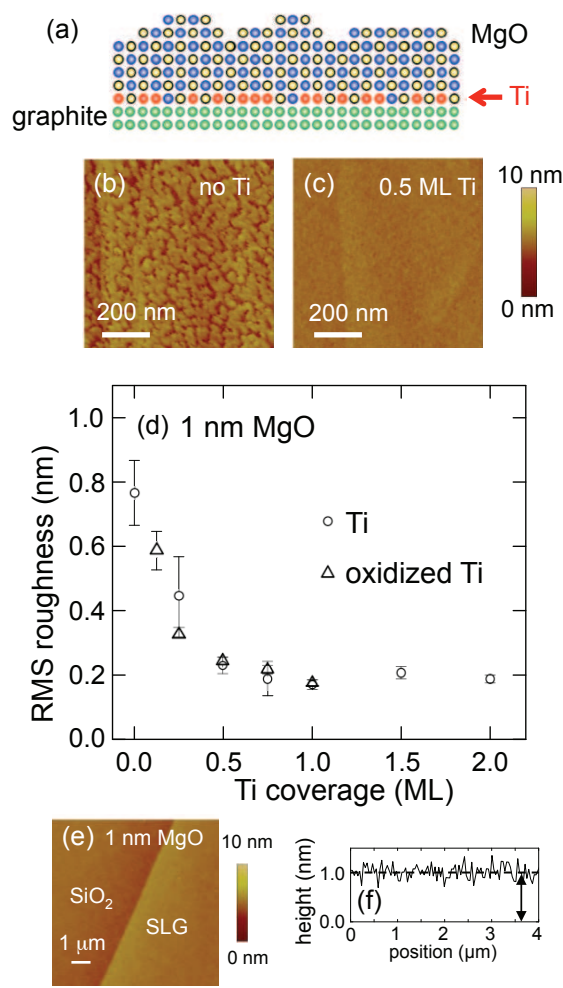

Figure 5. MgO growth on HOPG and SLG; (a) schematic of Ti seeding on HOPG or SLG prior MgO growth. (b) AFM image of 1 nm MgO without the Ti seeding. (c) AFM image of 1 nm MgO grown on 0.5 ML Ti/HOPG. (d) The rms roughness of 1 nm MgO films as a function of Ti coverage. Open circles are for samples in which the MgO is deposited immediately after Ti. Open triangles are for samples that are exposed to 30 L of $O_2$ gas to fully oxidize the Ti prior to MgO growth. (e) AFM image of 1 nm MgO grown by e-beam deposition in UHV on a graphene sheet dressed by 1.0 ML Ti (with post oxidation). There is no aggregation of MgO at the edges of the graphene. (f) Typical AFM line cut of 1nm MgO film on graphene surface.

Figure 6

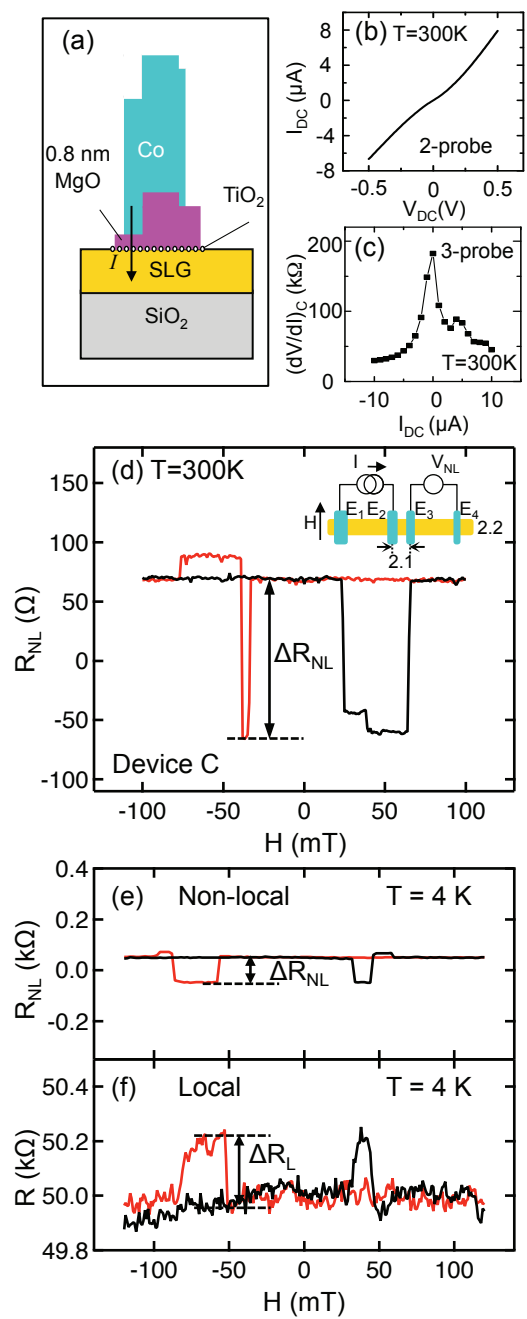

Figure 6. The SLG spin valve device with tunneling contacts. (a) Schematic drawing of the Co/MgO, $TiO_2$/SLG tunneling contacts. The arrow indicates the current flow. (b) Typical 2-probe IV curve between Co electrodes through SLG measured at 300 K. (c) Typical differential contact resistance $(dV/dI)_C$ as a function of bias current measured at 300 K. (d) Non-local MR scans of a SLG spin valves measured at room temperature (Device C, SLG, tunneling contacts, $L$ = 2.1 μm, $W$ = 2.2 μm, $V_g$ = 0 V). The non-local MR of 130 Ω is indicated by the arrow. Inset: the non-local spin transport measurement. (e-f) Nonlocal and local MR loop of device C measured at 4 K ($V_g$ = 0 V).

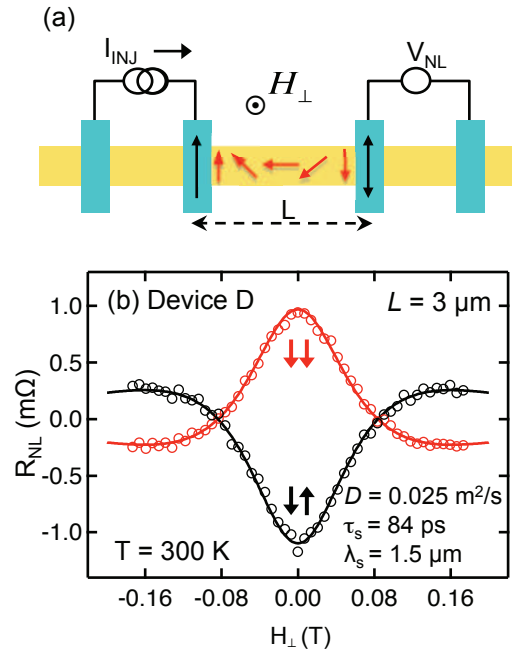

Figure 7. Hanle spin precession in SLG spin valves. (a) Schematic drawing of Hanle measurement by applying a out of plane magnetic field. (b) Non-local resistance as a function of the out-of-plane magnetic field of device D at 300 K (SLG, transparent contacts, $L = 3$ μm, $W \sim 2$ μm, $V_g = 0$ V). The red (black) circles are data for parallel (antiparallel) alignment of the central electrodes. The red and black lines are curve fits based equation 2.

Figure 8

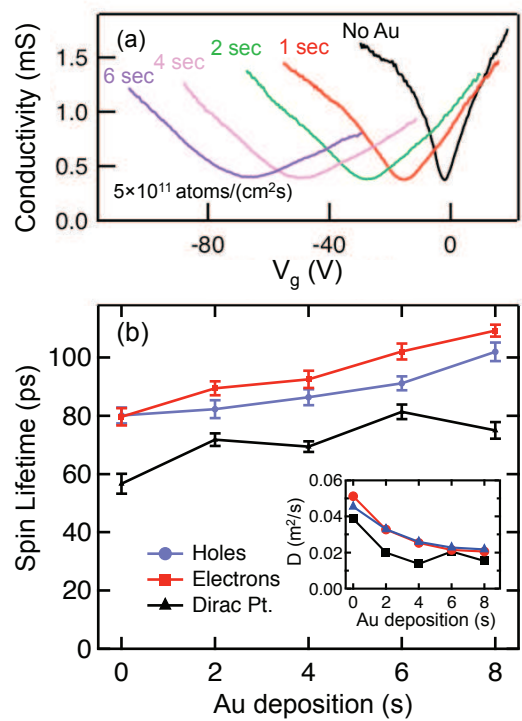

Figure 8. Electrical and spintronic properties of SLG by Au doping at 18 K (Device E, SLG, transparent contacts, $L = 2.5$ μm). (a) the gate dependent conductivity at selected values of Au deposition. (b) spin lifetime of SLG as a function of the Au deposition for holes with concentration of $2.9 \times 10^{12}$ cm$^{-2}$ (blue circles), for electrons with concentration of $2.9 \times 10^{12}$ cm$^{-2}$ (red squares), and at the Dirac point (black triangles). Error bars represent the 99% confidence interval. Inset: the diffusion constant as a function of Au deposition. Error bars are omitted when they are comparable to the symbol size.

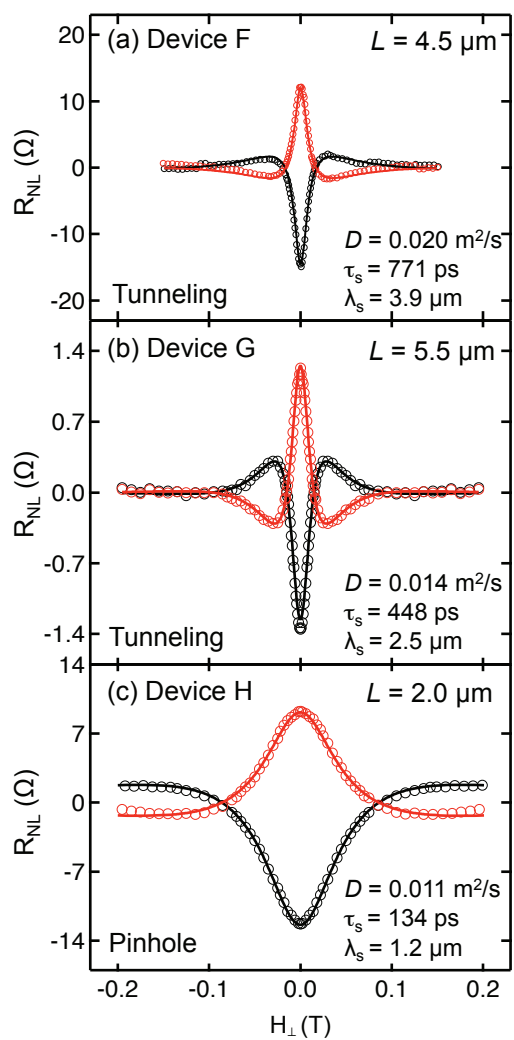

Figure 9. Hanle spin precession for SLG spin valves with different contacts at 300 K. (a) with tunneling contacts for $L = 4.5$ μm for $V_g = V_D$ (Device F, SLG, tunneling contacts). (b) with tunneling contacts for $L = 5.5$ μm for $V_g = V_D$ (Device G, SLG, tunneling contacts). (c) with pinhole contacts for $L = 2.0$ μm for $V_g = V_D$ (Device H, SLG, pinhole contacts). The top (red) and bottom (black) curves correspond to Hanle curves of the parallel and anti-parallel states, respectively. The solid lines are best fit curves based on equation 2.

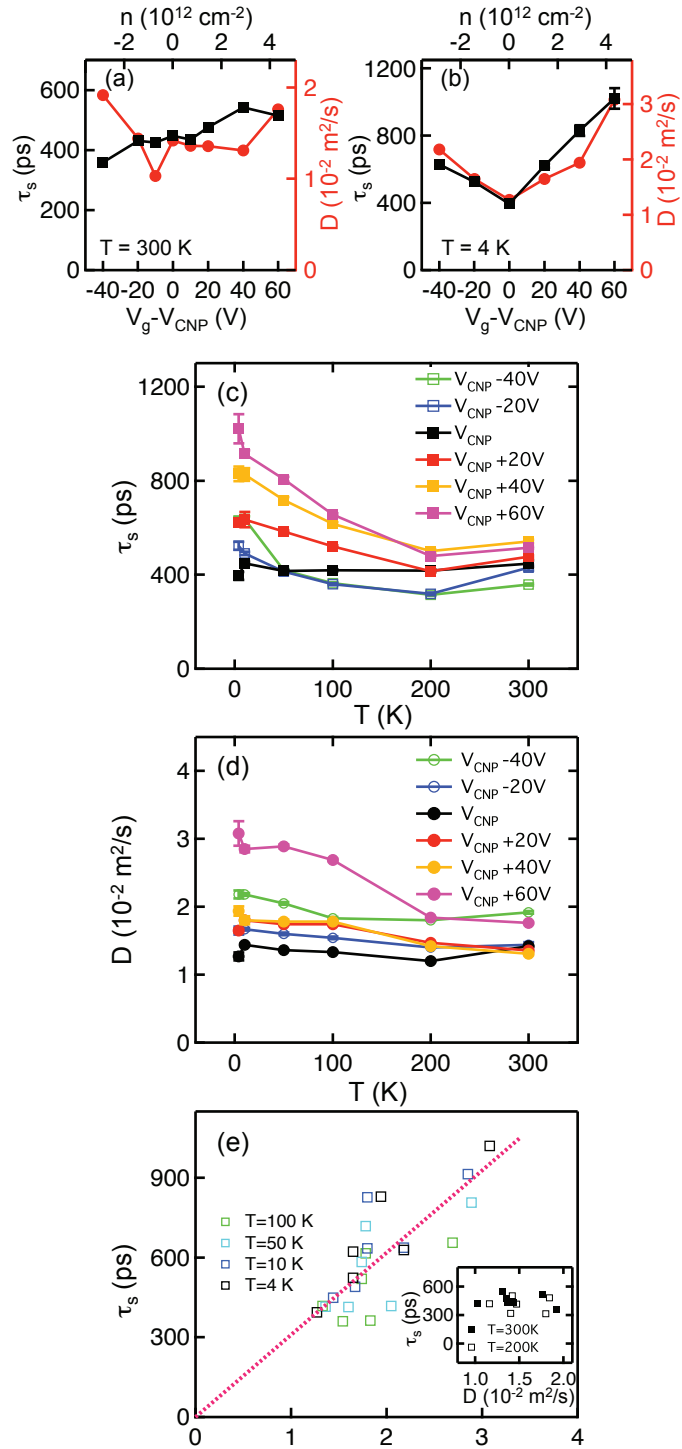

Figure 10. Spin relaxation in SLG (Device G, SLG, tunneling contacts, $L = 5.5$ μm). (a-b) spin lifetime (squares) and diffusion coefficient (circles) as a function of gate voltage at 300 K and 4 K. (c-d) temperature dependence of spin lifetime and spin diffusion coefficient at different gate voltages relative to the charge neutrality point. (c) Plot of spin lifetime vs. diffusion coefficient at T ≤ 100 K. The dotted line is a linear fit of the spin lifetime vs. diffusion coefficient. Inset: plot of spin lifetime vs. diffusion coefficient at T >100 K.

Figure 11

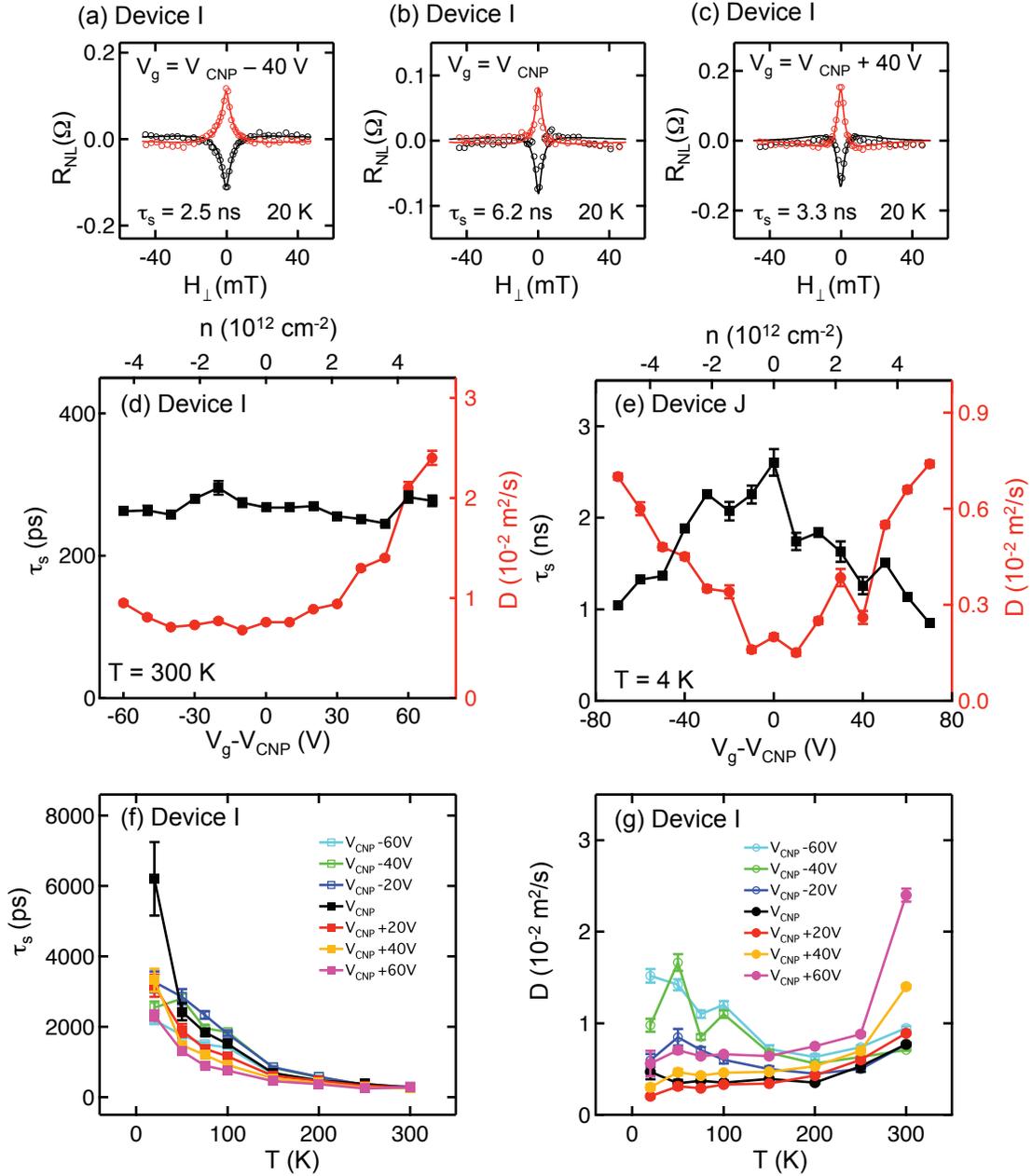

Figure 11. Spin relaxation in BLG (Device I and J). (a-c) Hanle measurement at 20 K for $V_g - V_{CNP}$ = –40 V, 0 V, and +40 V, respectively (Device I, BLG, tunneling contacts, $L = 3.1$ μm). (d) Spin lifetime (black squares) and diffusion coefficient (red circles) as a function of gate voltage at 300 K (Device I). (e) Spin lifetime (black squares) and diffusion coefficient (red circles) as a function of gate voltage at 4 K (Device J, BLG, tunneling contacts, $L = 4.0$ μm). (f-g) Temperature dependence of spin lifetime and spin diffusion coefficient at different gate voltages (Device I).

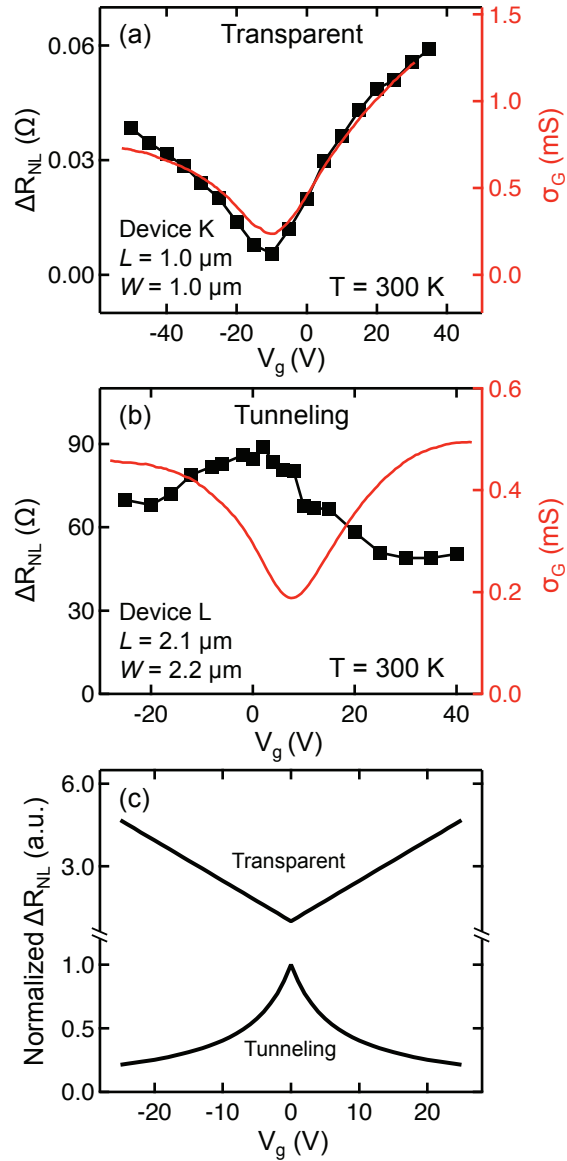

Figure 12. Gate tunable spin transport in SLG spin valves. (a) non-local MR (black squares) and conductivity (red lines) as a function of gate voltage for device K at 300 K (SLG, transparent contacts). (b) (a) non-local MR (black squares) and conductivity (red lines) as a function of gate voltage for device L at 300 K (SLG, tunneling contacts). (c) numerical prediction of the non-local MR as a function of gate voltage for transparent and tunneling contacts (the curves are normalized by their value at zero gate voltage).

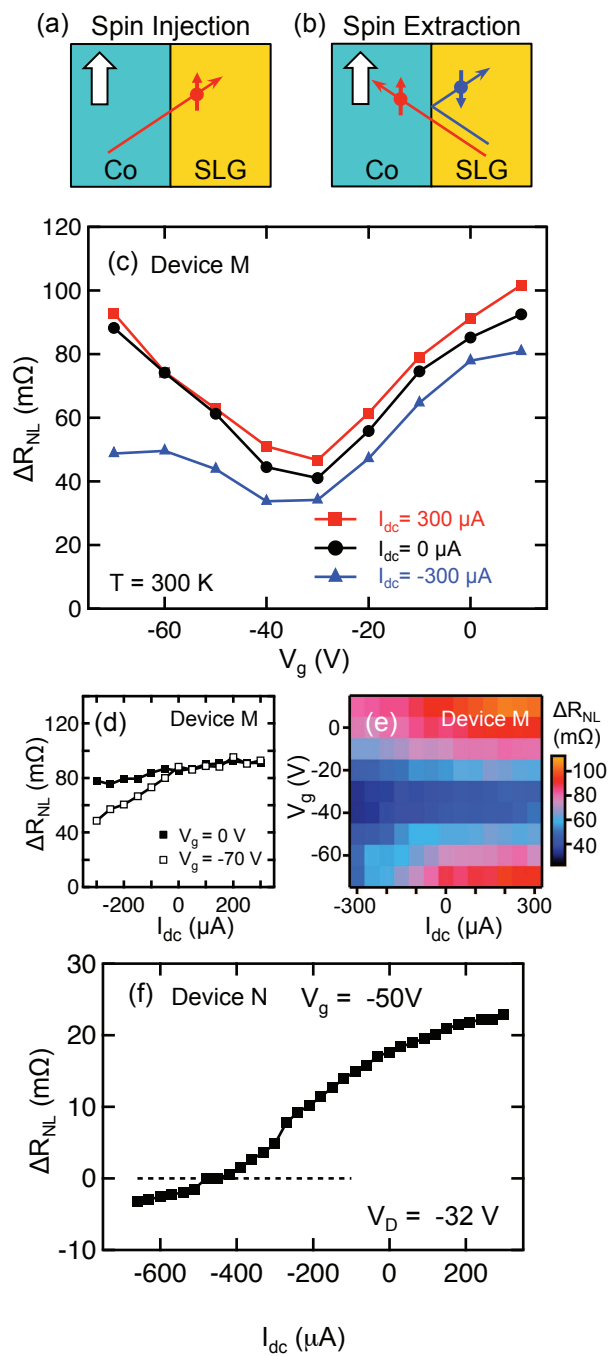

Figure 13. Bipolar spin transport in SLG spin valves with transparent contacts at 300 K (Device M, $V_D = -34$ V, and device N, $V_D = -32$ V). (a-b) schematic drawing of spin injection under positive bias and spin extraction under negative dc bias current. (c) non-local MR as a function of gate voltage for device M at $I_{dc}= 0$ μA, $I_{dc}=300$ μA, and $I_{dc}=-300$ μA. (d) non-local MR as a function of dc bias current for device M at $V_g = 0$ V (electrons, solid squares) and -70 V (holes, open squares). (e) nonlocal MR as a function of gate voltage and dc bias current for sample A. (f) nonlocal MR as a function of dc bias current at the hole side for device N. The dashed line shows the zero value.